\documentclass[sigconf]{acmart}

\AtBeginDocument{%
  \providecommand\BibTeX{{%
    \normalfont B\kern-0.5em{\scshape i\kern-0.25em b}\kern-0.8em\TeX}}}

\setcopyright{acmcopyright}
\copyrightyear{2024}
\acmYear{2024}
\acmDOI{xx.xxx/xxx_x}
\acmConference[]{}{}{}
\acmPrice{}
\acmISBN{979-8-4007-1637-9}

\usepackage{graphicx}
\usepackage{tabularx} 
\usepackage{booktabs}
\usepackage{amsmath}
\usepackage{caption}
\usepackage{multirow}
\usepackage{float}
\usepackage{afterpage}

\copyrightyear{2024}
\acmYear{2024}
\setcopyright{rightsretained}
\acmConference[ICMLT 2024]{2024 9th International Conference on Machine Learning Technologies (ICMLT)}{May 24--26, 2024}{Oslo, Norway}
\acmBooktitle{2024 9th International Conference on Machine Learning Technologies (ICMLT) (ICMLT 2024), May 24--26, 2024, Oslo, Norway}\acmDOI{10.1145/3674029.3674070}
\acmISBN{979-8-4007-1637-9/24/05}

\begin{document}

\title{A GNN Model with Adaptive Weights for Session-Based Recommendation Systems}

\author{Begüm Özbay}
\affiliation{%
  \institution{Istanbul Technical University,\\
Department of Computer\\Engineering\\}
  \streetaddress{1 Th{\o}rv{\"a}ld Circle}
  \city{Istanbul}
  \country{Turkey}}
\email{ozbaybe21@itu.edu.tr}

\author{Resul Tugay}
\affiliation{%
  \institution{Gazi University,\\
Department of Computer\\Engineering\\}
  \streetaddress{8600 Datapoint Drive}
  \city{Ankara}
  \country{Turkey}}
\email{resultugay@gazi.edu.tr}

\author{Şule Gündüz Öğüdücü}
\affiliation{%
  \institution{Istanbul Technical University,\\
Department of Artificial Intelligence\\ and Data Engineering\\}
  \streetaddress{1 Th{\o}rv{\"a}ld Circle}
  \city{Istanbul}
  \country{Turkey}}
\email{sgunduz@itu.edu.tr}

\renewcommand{\shortauthors}{B. Özbay, et al.}

\begin{abstract}
  Session-based recommendation systems aim to model users' interests based on their sequential interactions to predict the next item in an ongoing session.  In this work, we present a novel approach that can be used in session-based recommendations (SBRs). Our goal is to enhance the prediction accuracy of an existing session-based recommendation model, the SR-GNN model, by introducing an adaptive weighting mechanism applied to the graph neural network (GNN) vectors. This mechanism is designed to incorporate various types of side information obtained through different methods during the study. Items are assigned varying degrees of importance within each session as a result of the weighting mechanism. We hypothesize that this adaptive weighting strategy will contribute to more accurate predictions and thus improve the overall performance of SBRs in different scenarios. The adaptive weighting strategy can be utilized to address the cold start problem in SBRs by dynamically adjusting the importance of items in each session, thus providing better recommendations in cold start situations, such as for new users or newly added items. Our experimental evaluations on the Dressipi dataset demonstrate the effectiveness of the proposed approach compared to traditional models in enhancing the user experience and highlighting its potential to optimize the recommendation results in real-world applications.
\end{abstract}



\maketitle

\section{Introduction}

Recommender systems (RSs) have been playing an increasingly important role in e-commerce. In recent years, these systems have demonstrated significant effectiveness in delivering personalized suggestions to users, particularly through the utilization of deep learning techniques. In contrast to recommender systems that rely solely on user-item interactions without considering sessions, session-based recommendation systems predict user preferences or actions within a single interaction session rather than over a longer period. By analyzing the sequence of user interactions like clicks or views, these systems recommend items that are expected to be similar to the items within that particular session. Session-based recommender systems outperform traditional ones such as collaborative filtering and content-based filtering, particularly when users' past preferences are not available and change over time. Additionally, these systems have gained great popularity recently due to their ability to detect short-term or contextual user preferences for items. However, there are two challenges in session-based recommendation systems. Firstly, due to user anonymity, personal data cannot be utilized for recommendations. Secondly, there's variability in session lengths and the shortness of the sessions, especially in real-time data.

Session-based recommendation systems are important in adapting to the dynamic nature of user interactions in online platforms, enabling personalized recommendations aligned with users' session-specific behaviors. Despite the advances brought by the SR-GNN (Session-based Recommendation with Graph Neural Network) \cite{SRGNN} model in capturing sequential patterns and temporal dynamics in user sessions, a significant challenge remains in adapting to the varying importance levels of different elements within a session. This limitation allowed the development of an improved SR-GNN model that includes adaptive weights to dynamically adjust the importance of elements based on their relevance to the ongoing session. This development marks an important step towards improving the accuracy and personalization of recommendations within session-based recommendation systems.

Integration of adaptive weighting mechanisms within recommendation models allows for a more accurate assessment of element significance across user sessions. Traditional RSs often struggle to accurately reflect users' evolving preferences throughout a session. In response to this challenge, our proposed model with adaptive weights aims to address the limitations inherent in traditional approaches by dynamically adapting to the importance of items in each session. In this paper, we explore the technical details of the enhanced GNN model, summarizing its novel features, and presenting compelling experimental results on the Dressipi\cite{dressipi} dataset. These results demonstrate the effectiveness of adaptive weights in significantly improving prediction accuracy and suitability for session-based recommendation systems.

Thorough comparative analyses were carefully conducted to evaluate the performance of the enhanced SR-GNN model with adaptive weights. Our experimentation encompassed the consideration of several baseline models, notably including SR-GNN, NARM\cite{NARM}, and TAGNN\cite{TAGNN}, to establish a robust benchmark for comparison. Additionally, we introduced variations to the SR-GNN base model by incorporating side information, which encompasses supplementary data or features aimed at enriching the model's understanding or improving its predictive capabilities. Two distinct strategies were employed: firstly, the inclusion of side information related to the last item in the session, enhancing local representation; secondly, the incorporation of side information for all items in the session averaged to provide a global representation. The results from the diverse set of experiments provided valuable insights into the model's performance under various configurations. The baseline models served as crucial references for evaluating the efficacy of our proposed enhancements. Notably, the SR-GNN model augmented with adaptive weights consistently outperformed these baselines, demonstrating significant improvements in recommendation accuracy and relevance.

\section{Related Work}
In this section, we first summarize deep learning applications for a session-based recommendation and then review the applications that incorporate side information to improve item prediction performance.

Deep Learning-based SBRs have gained importance recently due to their effectiveness in modeling the relationship between items in users' shopping history.  This approach allows researchers to develop various models for SBRs. GRU4Rec\cite{GRU4REC} adopts a gated recurrent unit (GRU) to capture the sequential behavior of users. Self-attention networks were also introduced in SBRs due to the significant performance improvement in the natural language processing (NLP) area. NARM employs an attention mechanism in RNN to capture the relationships in the sequence, and then generate the final representation of the session. TAGNN uses a target attention mechanism to capture the attention relationship between all session items and target items. TAGNN++\cite{TAGNN++} applies self-attention to TAGNN with the transformer module to learn a richer representation. SASRec\cite{SASREC} proposes a self-attention-based sequential model. Like SASRec, Transformer and BERT\cite{BERT} were used to substitute attention representation modules in Transformer4Rec\cite{Transformer4Rec} and BERT4Rec\cite{BERT4REC} respectively, which benefit from pre-trained models using self-supervised learning and show remarkable performance in the next item prediction task. GNN gives a new perspective for SBRs, which takes the items in sessions as graph nodes and models complex dependency among items instead of a simple left-to-right one-directional sequence. SR-GNN models all session items into a session graph and uses a GNN to learn the transition relationship. NISER\cite{NİSER} uses normalized representations to help better learn long tail items and less popular new items. MSGIFSR\cite{MSGIFSR} captures the interaction between different granularity intent units and relieves the difficulty of capturing long-range dependency, thus decreasing the information loss.

NirGNN\cite{NirGNN} is a novel GNN model for session new item recommendation (GSNIR) that addresses the challenge of recommending items that users have never interacted with in past sessions. To overcome this problem, NirGNN introduces a dual-intent learning method to simulate user decision-making processes and uses a zero-shot learning approach to extract new item representations. SimCGNN\cite{SimCGNN} introduces a model designed to solve the same last-item confusion problem by using the contrastive module to increase the differentiation between sessions interacting with the same last item.

Three prominent models, SR-GNN, NARM, and TAGNN, were selected as baseline models in our study on session-based recommendation tasks using the Dressipi dataset. 

SR-GNN, serving as the foundational inspiration for the proposed new model, pioneers the modeling of sessions as graphs. It employs a GNN to create session graphs, obtaining item representations through a combination of global and local session representations processed by an attention network. The recommendation score is derived by multiplying the candidate item embedding with the relevant session representation, and SoftMax activation function is applied to represent the probability of the next click. The model is trained through cross-entropy loss minimization.

NARM, designed to capture both short-term and long-term user interests, adopts a hybrid encoder to summarize the entire session and selectively focus on important items. It calculates the similarity between the resulting session representation and item embeddings, followed by SoftMax and ranking to derive recommendation scores. This approach effectively addresses the challenge of modeling sequential patterns in user behavior, resulting in more accurate recommendations.

TAGNN introduces a target-aware attention mechanism that applies weighting by comparing the session representation with target item representations. This ensures that the model reflects the learned level of interest for the target item to be recommended. The adaptive activation feature allows TAGNN to create custom user interest profiles for different target items, enhancing the personalization and effectiveness of recommendations.

In contrast, we propose a novel framework that learns item weights via GNN, thereby assisting deep-learning models in achieving better learning outcomes. Additionally, prior deep learning models can integrate our weighting mechanism into their models.

However, these deep learning-based methods only consider item-level information and ignore metadata information, which is essential for items that occur in historical sessions. The utilization of SBRs with side information in recommendation models is also an important research area. These models can provide more personalized recommendations using side information such as item features, users' background information, or other personal characteristics. FDSA\cite{FDSA} integrates item-level transitions and feature-level transitions by adding a self-attention mechanism based on SASRec. IAGNN\cite{IAGNN} creates a category-sensitive graph with both item and item category nodes to represent complex transitions in the session. CoHHN\cite{CoHHN} models price and interest preference simultaneously with the heterogeneous hypergraph network for SBRs. With DIF-SR\cite{DIF-SR}, side information is moved from the input to the attention layer, providing higher flexibility for attention representation. SIHG4SR\cite{SIHG4SR} uses heterogeneous graphs to represent the items view sequences and the item categories about the items together. Heterogeneous GAT is used to model the complex relationship in this type of graph structure.

The proposed methodology enables significant improvements in session-based recommendation systems because the integration of item features with the developed weighting mechanism provides more personalized recommendations by better aligning with users' interests. This approach improves the accuracy of recommendations and user experience. By incorporating item attributes, the recommendation system gains a deeper understanding of each item's characteristics, allowing for more precise matching with users' preferences. Furthermore, the weighting mechanism enhances the relevance of recommendations by prioritizing items that are more likely to resonate with users.

In conclusion, MRR@20 and Recall@20 are invaluable metrics for evaluating the performance of recommender systems, providing insights into their ability to generate relevant and well-ranked recommendations. These metrics serve as essential tools for assessing the effectiveness of recommendation algorithms in adapting to user preferences and optimizing user experience in various application domains.

\section{Methodology}
In this section, we provide an overview of the base model, SR-GNN, and explain the integration of the devised adaptive weighting method along with item side information into the model architecture.
\subsection{Background Method}
SR-GNN is a leading model in session-based recommendation systems that utilizes a GNN. This architecture effectively represents transitions between items during sessions, allowing it to capture complex interdependencies among items. By identifying underlying trends in user behavior, the algorithm generates accurate predictions for next-item recommendations. Leveraging graph topologies, SR-GNN significantly enhances both the scalability and accuracy of existing session-based recommendation algorithms.

\subsection{Proposed Method}
The objective of session-based recommendation is to predict the subsequent item a user will click, relying solely on the user's current sequential session data, without access to their long-term preference profile. In the context of a session-based recommendation system, a collection of items list represented with $\mathcal{V} = \{v_1, v_2, \ldots, v_{|\mathcal{V}|}\}$, and a session sequence represented with $S_i = [v^{(i)}_1, v^{(i)}_2, \ldots, v^{(i)}_{N}]$, where each $v^{ (i)}_t$  is an element from $\mathcal{V}$ included in a given session $i$. The number of interactions within the session is denoted by $N$ and referred to as session length.

\begin{figure*}[htbp]
  \centering
  \includegraphics[width=1\linewidth]{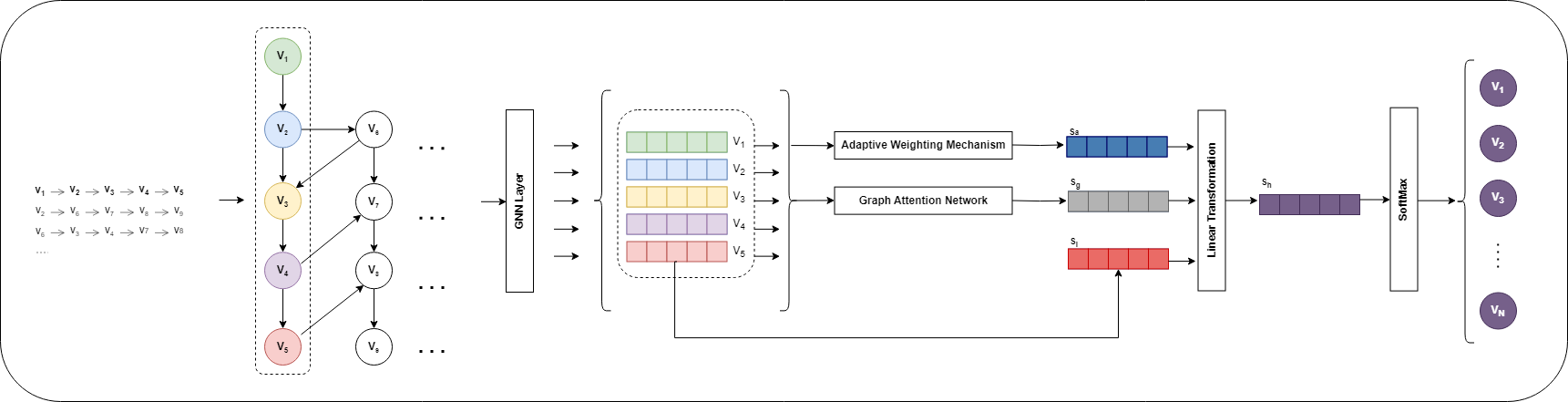} 
  \caption{The workflow of the proposed method.}
  \label{fig: method}
\end{figure*}

Figure \ref{fig: method} illustrates the workflow of the proposed method. The model consists of four steps: modeling sessions as graphs, learning node representations, generating session representations, and making recommendations. The steps of modeling sessions as graphs and learning node representations are based on methods applied in the SR-GNN method. In the step of generating session representations, in addition to the global preference $Sg$ and the current interest $Sl$ of the user in that session, the $Sa$ representation obtained with the session items weighted by the applied weighting mechanism is included. After obtaining the session representation ($Sa$) with the session items weighted by the applied weighting mechanism, we combine it with each session's global preference ($Sg$) and the user's current interest ($Sl$) in that session through a simple linear transformation to obtain the hybrid embedding. Within the framework of a session-based recommendation model, we generate a probability vector for every possible item. Each element within the vector represents the recommendation score of the corresponding item. The items with the high scores are considered potential recommendations for the session. We demonstrate the effectiveness of the proposed method through extensive experiments conducted on real-world datasets.

At first, to better predict users' next clicks, an approach was devised in which the side information associated with viewed items within a session was incorporated. This approach consists of two distinct stages and leverages side information representations to enrich the inherent characteristics of item representations. Firstly, the side information corresponding to the last item viewed within the session was appended to the local representation of the item. This method, termed SR-GNN-SI, aimed to capitalize on the most recent user interest to inform subsequent predictions. In this process, each item is represented as a vector in a latent space with its features. Adding side information to the last viewed item is intended to improve understanding of user preferences in the context of the session. Secondly, a complementary strategy was implemented whereby the session mean of side information was computed by aggregating the side information vectors of all items viewed within the session, and subsequently added to the representation vectors. This representation, referred to as the SR-GNN-MSI method, aimed to encapsulate the features of all viewed items within the session. Through the averaging process, the session-level representation, which serves as a global representation, was derived.

Session ends play a critical role in understanding changing trends that reflect current user preferences and consider short-term preference changes. It is extremely important to understand these changing trends over time in sessions. In our work, we develop a custom weighting formula that dynamically evaluates the user's interactions over time within the session to focus on the final parts of the session and prioritize session elements. Using the session element weights calculated in this formula, we create a strategy that allows it to better understand current preferences. This weighting strategy is designed to prioritize some parts of the session, in line with our goal of providing users with more effective and personalized recommendations.

\begin{figure}[htbp]
  \centering
  \includegraphics[width=1\linewidth,height=4cm]{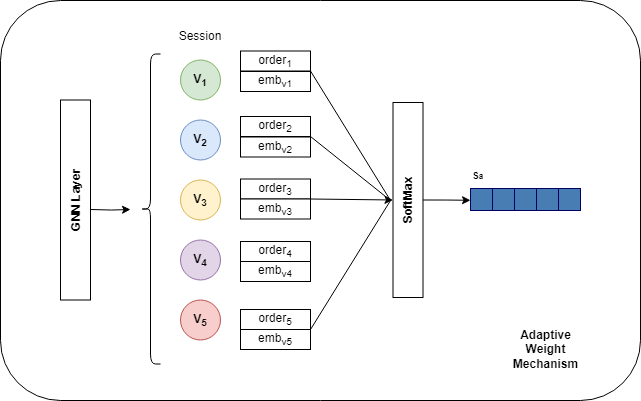} 
  \caption{The adaptive weighting mechanism.}
  \label{fig: adaptive weighting mechanism}
\end{figure}

The vectors derived from the GNN are weighted to incorporate an adaptive feature. When the similarity results between the vector of each session element and the last element are passed directly to the SoftMax layer, the element with the highest similarity value will have the highest weight. However, considering that the closeness of an item to the end of a session is supposed to enhance its significance within the session, we decided to incorporate session order information as well. We assign a new order to the items using the formula in Equation (2). Cosine similarity and new orders together are used to determine item weights for the session. A workflow diagram for adaptive weighting is shown in Figure \ref{fig: adaptive weighting mechanism}.

Each session $s$ can be represented by an embedding vector $\mathbf{s}$, which is composed of node vectors learned via graph neural networks. Let
$[\mathbf{v}_{1}^{i}, \mathbf{v}_{2}^{i}, \ldots, \mathbf{v}_n^{i}]$ be the list of node vectors in session $s$.  For session $i$, cosine similarity is calculated between each element $\mathbf{v}^{(i)}_1, \mathbf{v}^{(i)}_2, \ldots, \mathbf{v}^{(i)}_{n-1}$ and the last element. Here, $\mathbf{v}^{(i)}_{\text{n}}$ represents the vector of the last element and $\mathbf{v}^{(i)}_j$ represents the vector of each element in the session.

\begin{equation}
\label{eq: cosine}
cos = (\mathbf{v}^{(i)}_j, \mathbf{v}^{(i)}_{n})
\end{equation}
\begin{equation}
\label{eq:newOrder}
new\_orders = \exp^{(i/t)}
\end{equation}

\begin{equation}
\label{eq:Weights}
weights = softmax(\cos*new\_orders)
\end{equation}

When calculating the similarity between all items in the session and the last item, we derived a formula that can use the similarity between items and the order information of the items in the session, based on the assumption that the importance of the item increases with its proximity to the end of the session. To describe the relationship between cosine similarity and order, we set a parameter $t$. This parameter controls the order of each item. In situations where similarity is prioritized over item order, it is advisable to increase the parameter $t$. Conversely, when order significance is increased, a reduction in $t$ is recommended.
We consider equal importance on both the sequencing of items within a session and the effects of the computed similarity values in the weight calculation.

\section{Experimental Results}
In this section, we first evaluate the performance of three distinct baseline models against the proposed model. Then, in the second set of experiments, we further evaluate the enhancements of each SR-GNN model by incorporating both the last item side information and mean side information. These enhancements were applied at varying data rates, denoted by the parameter $1/k$. The $1/k$ notation represents the fraction of the original dataset used for training, with $k$ being an integer. For instance, if k is 64, then 1/64 of the dataset was utilized for training. By varying the parameter $k$, different proportions of the dataset were sampled for training, allowing for an analysis of model performance under varying data availability conditions.

In another experiment, adaptive weights were introduced to address the cold start problem, which refers to the challenge of making accurate predictions for new or rarely seen items with limited available data. The notation $1/k$ denotes the fraction of the dataset used for training, where a smaller value of $k$ indicates a smaller proportion of the dataset being utilized. By using a smaller portion of the dataset, the cold start problem arises as less historical data is available for training models, making it more challenging to generate accurate predictions for items that have sparse interactions in the dataset.
Additionally, parameters such as how long the sessions would be and what the $t$ value would be when using adaptive weight were also examined. These experiments aimed to evaluate the performance of different model configurations and to determine the most effective parameters that will provide a solution to the cold start problem.
\subsection{Dataset}

In this study, we analyze Dressipi's public dataset from the 2022 RecSys challenge, which contains one million online retail sessions based on item sequences and corresponding label data. All items in this dataset are tagged with content data, and tags are assigned using Dressipi's human-in-the-loop system. The dataset was sampled and anonymized. Since sessions are usually short, it was emphasized that it is important to have a high-fidelity in-session recommender that can respond to the user's current session activities and make recommendations that can create the best possible experience. The Dressipi dataset is focused on buying sessions, with each session containing one item purchased. 

The total number of items in the session data, the features of the items in the label data, and the different values of these features are given in Table \ref{tab: Dressipi dataset}. Items have 904 unique feature-value pairs, such as the pair 'color: green'. In addition, each item has a different number of features because the types of items vary in the data, which includes clothing and footwear.

\begin{table}[htbp]
  \centering
  \caption{Dressipi dataset.}
  \begin{tabular}{lcr}
    \toprule
    Description & Value \\
    \midrule
    Num. items   & 23691   \\
    Num. item features   & 73   \\
    Num. feature-value pairs   & 904   \\
    Avg. session length   & 4.74   \\
    Avg. features of an item   & 19.47   \\
    \bottomrule
  \end{tabular}
  \label{tab: Dressipi dataset}
\end{table}

In neural network approaches used as recommendation models for session-based recommender systems, it has been shown that training on a more recent part of the dataset yields slightly better results than training on the entire dataset. Since we are working with data that is large and needs to account for user behavior that changes over time, three different data sets obtained by proportioning were used in the experiments.
The 17-month dataset was first sorted chronologically, and the sessions of the subsequent day were used as testing sessions. Then, we reported our results on models trained on the more recent 1/128, 1/64, 1/32, and 1/4 fractions of the training sequences and the full data. Note that when we train the model only on the more recent fractions, some items in the test set will not appear in the training set. 

This means that some items in the test set are not seen and learned by the model, especially if older data fractions are not used. 

This can create a training case that is closer to the real-world use of the model since in the real world there are user preferences and behaviors that change over time. Therefore, focusing the model on more recent data can improve recommendation performance, but it can also result in some elements being missing. Being aware of this is important when evaluating the performance of the model. 

The statistics of the different datasets are shown in Table \ref{tab: dataset statistics}.

\begin{table}[h]
\centering
\caption{Statistics of the datasets used in our experiments.}
\begin{tabular}{lccc}
  \toprule
  Datasets & Sessions & Unique Items & Avg. Length \\
  \midrule
  Dressipi 1/128 &  28073 & 3309 & 8.8 \\
  Dressipi 1/64 & 58244 & 4188 & 9.2 \\
  Dressipi 1/32 &  112295 & 4651 & 8.7 \\
  Dressipi 1/4   &  931919 & 8641 & 7.4 \\
  Dressipi 1/1   &  3727678 & 22549 & 7.3 \\
  \bottomrule
\end{tabular}
\label{tab: dataset statistics}
\end{table}

\subsection{Performance Evaluation}
Since recommender systems can only recommend a few items at a time, the actual item that the user can select should be among the first few items on the list. We therefore use the following two key metrics to evaluate the performance of recommendation lists. These metrics are commonly used to evaluate how effectively recommendation systems adapt to real-world use cases.
\begin{itemize}
    \item MRR@20: Mean Reciprocal Rank (MRR), which is the average of reciprocal ranks of the desired items. In the case of a rank higher than 20, the reciprocal rank is set to zero. MRR considers the ranking of each item, in scenarios where the sequence of recommendations is important.
    \item Recall@20: The Recall@20 is the proportion of cases in which the desired item appears among the top 20 items. This metric plays a critical role in determining the effectiveness of recommendation systems. 
\end{itemize}

\subsection{Results}

Base models were chosen due to their established performance in the field of session-based recommendation, which allows easy evaluation of proposed improvements. We used the 1/64, 1/4, and 1/1 ratios as inputs for the SR-GNN, NARM, and TAGNN base models, and the results are included in Table \ref{tab: baseline results}. These ratios indicate different levels of data representation; where 1/1 corresponds to the entire dataset, 1/64 and 1/4 represent retrieving the last parts when the sessions are sorted.

\begin{table}[htbp]
  \centering
  \captionof{table}{Results for baselines.}
  \begin{tabular}{lcr}
    \toprule
        \multirow{2}{*}{Method} & \multicolumn{2}{c}{\textbf{Performance Metrics}} \\
    \cmidrule{2-3}
    & \textbf{Recall@20} & \textbf{MRR@20} \\
    \midrule
    SR-GNN (1/64)  & 33.54   & 11.73   \\
    NARM (1/64) & 45.01   & 15.04   \\
    TAGNN (1/64) & 39.75   & 17.10   \\
    \midrule
    SR-GNN (1/4)  & 44.75   & 16.26   \\
    NARM (1/4)  & 46.39   & 16.99   \\
    TAGNN (1/4)  & 51.05   & 17.45   \\
    \bottomrule
  \end{tabular}
  \label{tab: baseline results}
\end{table}

To enhance the model's capabilities, we introduced two variations: SR-GNN-SI and SR-GNN-MSI. SR-GNN-SI integrates side information from the last item into the local representation, while SR-GNN-MSI incorporates the average side information from all items in a session into the global representation. When the results are examined, it is seen that SR-GNN-MSI performs best on Recall@20 and MRR@20 metrics. In particular, SR-GNN-MSI shows higher performance than other configurations at 1/4 and 1/1 data representation ratios. This indicates that taking global structures of session data into account increases the predictive ability of the model. While SR-GNN-SI generally performs slightly better than SR-GNN, it provides a significant improvement in the Recall@20 metric for SR-GNN (1/4). However, for SR-GNN (1/64), this improvement is less pronounced. The results highlight the importance of using side information for session-based recommendation systems, especially global side information integration such as SR-GNN-MSI. This indicates that modeling user behavior more comprehensively can improve recommendation accuracy as shown in Table \ref{tab: results SI and MSI}.

\begin{table}[htbp]
  \centering
  \caption{Results for SI and MSI}
  \begin{tabular}{lcc}
    \toprule
        \multirow{2}{*}{Method} & \multicolumn{2}{c}{\textbf{Performance Metrics}} \\
    \cmidrule{2-3}
    & \textbf{Recall@20} & \textbf{MRR@20} \\
    \midrule
    \textbf{SR-GNN (1/1)} & 44.86 & 16.30 \\
    SR-GNN-SI & \textbf{44.92} & \textbf{16.31} \\
    SR-GNN-MSI & 45.05 & 16.38 \\
    \midrule
    \textbf{SR-GNN (1/4)} & \textbf{44.75} & \textbf{16.26} \\
    SR-GNN-SI & 45.71 & 16.79 \\
    SR-GNN-MSI & 45.96 & 16.73 \\
    \midrule
    \textbf{SR-GNN (1/64)} & \textbf{33.54} & \textbf{11.73} \\
    SR-GNN-SI & 33.84 & 11.90 \\
    SR-GNN-MSI & 33.73 & 11.74 \\
    \bottomrule
  \end{tabular}
  \label{tab: results SI and MSI}
\end{table}

Furthermore, an adaptive weights feature was introduced during the GNN stage, allowing for dynamic weighting of the similarity between the vector representation of the last item in a session and other items. Parameterized by $t$ experiments were conducted with values ranging from 2 to 5. Notably, optimal results were observed at $t=4$ as shown in Table \ref{tab: results adaptive weihts}.

\begin{table}[htbp]
  \centering
  \caption{Results for Adaptive Weights.}
  \begin{tabular}{lccc}
    \toprule
        \multirow{2}{*}{Method} & \multicolumn{2}{c}{\textbf{Performance Metrics}} \\
    \cmidrule{2-3}
    & \textbf{Recall@20} & \textbf{MRR@20} \\
    \midrule
    \textbf{SR-GNN (1/4)} & \textbf{44.75} & \textbf{16.26} \\
    SR-GNN-AW & 42.44 & 15.86 \\
    \midrule
    \textbf{SR-GNN (1/32)} & \textbf{37.18} & \textbf{13.53} \\
    SR-GNN-AW & 38.59 & 14.38 \\
    \midrule
    \textbf{SR-GNN (1/64)} & \textbf{33.54} & \textbf{11.73} \\
    SR-GNN-AW) & 38.18 & 14.19 \\
    \midrule
    \textbf{SR-GNN (1/128)} & \textbf{20.95} & \textbf{6.19} \\
    SR-GNN-AW & 31.68 & 11.56 \\
    \bottomrule
  \end{tabular}
  \label{tab: results adaptive weihts}
\end{table}

The performance of adding adaptive weights to the model has been observed at different data rates. In cases, with a data ratio of 1/64, better results were obtained by adding adaptive weights. However, in cases with larger ratios, such as the 1/4 data, a slight decrease in performance was observed. This points to the 'cold start' problem, which occurs when the model fails to make effective predictions for rarely viewed items. A decrease in the number of sessions resulting in a particular item leads to the occurrence of the cold start problem. This is especially evident in the data at 1/128 and 1/64 ratios. The cold start problem arises from the model's inability to sufficiently learn various session representations ending with the relevant item. Adding adaptive weights is one way to address this problem by making the model perform better. We further evaluated adaptive weights in the rates of 1/128 and 1/32 to support our hypothesis with experiments. In evaluating the experimental results, we can conclude that the adaptive weights method scales the relationship between items, increases the effect of rare items, and allows the model to make better predictions.
\begin{table}[htbp]
  \centering
  \caption{Results for adaptive weights.}
  \begin{tabular}{lccc}
    \toprule
        \multirow{2}{*}{Method} & \multicolumn{2}{c}{\textbf{Performance Metrics}} \\
    \cmidrule{2-3}
    & \textbf{Recall@20} & \textbf{MRR@20} \\
    \midrule
    \textbf{SR-GNN (1/4)} & \textbf{44.75} & \textbf{16.26} \\
    SR-GNN (T=4 Adaptive Weights) & 42.44 & 15.86 \\
    \midrule
    \textbf{SR-GNN (1/32)} & \textbf{37.18} & \textbf{13.53} \\
    SR-GNN (T=4 Adaptive Weights) & 38.59 & 14.38 \\
    \midrule
    \textbf{SR-GNN (1/64)} & \textbf{33.54} & \textbf{11.73} \\
    SR-GNN (T=4 Adaptive Weights) & 38.18 & 14.19 \\
    \midrule
    \textbf{SR-GNN (1/128)} & \textbf{20.95} & \textbf{6.19} \\
    SR-GNN (T=4 Adaptive Weights) & 31.68 & 11.56 \\
    \bottomrule
  \end{tabular}
  \label{tab: results adaptive weihts}
\end{table}

We supported this assumption with several experiments and obtained the best results for parameters $t=4$ and $t=3$, respectively, for values of $t$ ranging from 1-10. Table \ref{tab: results adaptive weihts} illustrates model results obtained with $t=4$. Our method has been thoroughly tested in several scenarios, including varying data set ratios of 1/64 and 1/4, as well as different session lengths. Session length ranges from 1-100 and represents the number of items viewed in that session.

Considering this long-range in session lengths, a limitation has been introduced for longer sessions as session endings are more relevant to current user interest. A threshold value $p$ has been determined for this limitation. If more items than $p$ were displayed in a session, the last $p$ item of the session was taken and adaptive weights experiments were carried out over these limited sessions. This  $p$ value has been tested for 5, 10, 15, and 20.
Table \ref{tab: session lengths} illustrates the impact of different session lengths on the SR-GNN model’s performance. The results show that the last 5 and last 10 outperformed the others. By concentrating on the end of sessions, the adaptive weights technique used in the SR-GNN model produced a notable improvement under the 1/64 data ratio with $t=4$ and session length 10.

\begin{table}[H]
  \centering
  \caption{Adaptive Weights Results for different settings.}
  \begin{tabular}{lcc}
    \toprule
        \multirow{2}{*}{Method} & \multicolumn{2}{c}{\textbf{Performance Metrics}} \\
    \cmidrule{2-3}
    & \textbf{Recall@20} & \textbf{MRR@20} \\
    \midrule
    \textbf{SR-GNN (1/64) (Last 5)} & 38.18 & 14.19 \\
    \textbf{SR-GNN (1/64) (Last 10)} & 38.30 & 14.31 \\
    \textbf{SR-GNN (1/64) (Last 15)} & 38.08 & 14.27 \\
    \textbf{SR-GNN (1/64) (Last 20)} & 37.92 & 13.95 \\
    \midrule
    \textbf{SR-GNN (1/4) (Last 5)} & 42.44 & 15.86 \\
    \textbf{SR-GNN (1/4) (Last 10)} & 42.84 & 15.87 \\
    \textbf{SR-GNN (1/4) (Last 15)} & 42.40 & 15.72 \\
    \textbf{SR-GNN (1/4) (Last 20)} & 39.74 & 14.80 \\
    \bottomrule
  \end{tabular}
  \label{tab: session lengths}
\end{table}

The experimental results show that using adaptive weighting at a data size of 1/64 provides a significant improvement. This shows that the use of adaptive weights is beneficial to improve the performance of the model. The 13.88\% increase in the recall rate emphasizes the importance of this improvement. For 1/128 and 1/32 data, increases of 51.36\% and 3.78\% were observed, respectively. However, at 1/4 data size, a 5.15\% drop in recall rate was unexpectedly observed. 

The drop in recall rate at the 1/4 data size can be attributed to the longer sequences encountered at this data scale, posing challenges for the model's predictive capabilities. Furthermore, the model contribution becomes more prominent due to its adeptness and successful results in cold start items, thereby improving its overall efficacy. The significant improvement observed at 1/128 and 1/64 is due to the model correctly predicting the cold start items. By effectively addressing the challenges posed by cold start scenarios, the model is able to leverage limited data more efficiently. This emphasizes the critical role of cold start prediction techniques in improving the model's performance across varying data ratios. Additionally, the demonstrated success in handling cold start items highlights the model's adaptability and robustness, further affirming its suitability for real-world recommendation systems.

\section{Conclusion}

This study presents a novel approach for session-based recommendation. Introducing an adaptive weighting mechanism to the GNN vectors, with a focus on considering the last item in a session, has proven to be a novel and effective approach. By computing similarities between all items within a session and the concluding item, a weighted scheme is established, assigning varying degrees of importance to items. Experimental evaluations on the Dressipi dataset demonstrate the superiority of the adaptive weighting strategy over the conventional SR-GNN model, highlighting its potential to markedly enhance predictive accuracy in session-based recommendation scenarios. Additionally, experiments conducted on datasets of varying sizes offer a solution to the cold start problem, further enhancing the versatility and applicability of the proposed method while suggesting its relevance in real-world applications.

In future work, we plan to investigate the scalability of the proposed method to larger datasets and more complex recommendation scenarios, as this could provide valuable insights into its practical applicability. Additionally, we aim to explore the incorporation of additional contextual information beyond the last item in a session, as this has the potential to further enhance the model's performance and could lead to novel findings in session-based recommendation research. These directions offer promising avenues for advancing the effectiveness and applicability of our proposed approach in real-world settings.

\begin{acks}
Begüm Özbay would like to thank Google DeepMind scholarship program for their support.
\end{acks}

\bibliographystyle{ACM-Reference-Format}
\bibliography{references}


\begin{thebibliography}{19}


\ifx \showCODEN    \undefined \def \showCODEN     #1{\unskip}     \fi
\ifx \showDOI      \undefined \def \showDOI       #1{#1}\fi
\ifx \showISBNx    \undefined \def \showISBNx     #1{\unskip}     \fi
\ifx \showISBNxiii \undefined \def \showISBNxiii  #1{\unskip}     \fi
\ifx \showISSN     \undefined \def \showISSN      #1{\unskip}     \fi
\ifx \showLCCN     \undefined \def \showLCCN      #1{\unskip}     \fi
\ifx \shownote     \undefined \def \shownote      #1{#1}          \fi
\ifx \showarticletitle \undefined \def \showarticletitle #1{#1}   \fi
\ifx \showURL      \undefined \def \showURL       {\relax}        \fi
\providecommand\bibfield[2]{#2}
\providecommand\bibinfo[2]{#2}
\providecommand\natexlab[1]{#1}
\providecommand\showeprint[2][]{arXiv:#2}

\bibitem[Cao et~al\mbox{.}(2023)]%
        {SimCGNN}
\bibfield{author}{\bibinfo{person}{Yuan Cao}, \bibinfo{person}{Xudong Zhang}, \bibinfo{person}{Fan Zhang}, \bibinfo{person}{Feifei Kou}, \bibinfo{person}{Josiah Poon}, \bibinfo{person}{Xiongnan Jin}, \bibinfo{person}{Yongheng Wang}, {and} \bibinfo{person}{Jinpeng Chen}.} \bibinfo{year}{2023}\natexlab{}.
\newblock \showarticletitle{SimCGNN: Simple Contrastive Graph Neural Network for Session-based Recommendation}.
\newblock  (\bibinfo{date}{2} \bibinfo{year}{2023}).
\newblock
\urldef\tempurl%
\url{http://arxiv.org/abs/2302.03997}
\showURL{%
\tempurl}


\bibitem[Cui et~al\mbox{.}(2021)]%
        {IAGNN}
\bibfield{author}{\bibinfo{person}{Chuan Cui}, \bibinfo{person}{Qi Shen}, \bibinfo{person}{Shixuan Zhu}, \bibinfo{person}{Yitong Pang}, \bibinfo{person}{Yiming Zhang}, \bibinfo{person}{Hanning Gao}, {and} \bibinfo{person}{Zhihua Wei}.} \bibinfo{year}{2021}\natexlab{}.
\newblock \showarticletitle{Intention Adaptive Graph Neural Network for Category-aware Session-based Recommendation}.
\newblock  (\bibinfo{date}{12} \bibinfo{year}{2021}).
\newblock
\urldef\tempurl%
\url{http://arxiv.org/abs/2112.15352}
\showURL{%
\tempurl}


\bibitem[Devlin et~al\mbox{.}(2018)]%
        {BERT}
\bibfield{author}{\bibinfo{person}{Jacob Devlin}, \bibinfo{person}{Ming-Wei Chang}, \bibinfo{person}{Kenton Lee}, {and} \bibinfo{person}{Kristina Toutanova}.} \bibinfo{year}{2018}\natexlab{}.
\newblock \showarticletitle{BERT: Pre-training of Deep Bidirectional Transformers for Language Understanding}.
\newblock  (\bibinfo{date}{10} \bibinfo{year}{2018}).
\newblock
\urldef\tempurl%
\url{http://arxiv.org/abs/1810.04805}
\showURL{%
\tempurl}


\bibitem[Guo et~al\mbox{.}(2022)]%
        {MSGIFSR}
\bibfield{author}{\bibinfo{person}{Jiayan Guo}, \bibinfo{person}{Yaming Yang}, \bibinfo{person}{Xiangchen Song}, \bibinfo{person}{Yuan Zhang}, \bibinfo{person}{Yujing Wang}, \bibinfo{person}{Jing Bai}, {and} \bibinfo{person}{Yan Zhang}.} \bibinfo{year}{2022}\natexlab{}.
\newblock \showarticletitle{Learning multi-granularity consecutive user intent unit for session-based recommendation}.
\newblock \bibinfo{journal}{\emph{WSDM 2022 - Proceedings of the 15th ACM International Conference on Web Search and Data Mining}}, \bibinfo{pages}{343--352}.
\newblock
\showISBNx{9781450391320}
\urldef\tempurl%
\url{https://doi.org/10.1145/3488560.3498524}
\showDOI{\tempurl}


\bibitem[Gupta et~al\mbox{.}(2019)]%
        {NİSER}
\bibfield{author}{\bibinfo{person}{Priyanka Gupta}, \bibinfo{person}{Diksha Garg}, \bibinfo{person}{Pankaj Malhotra}, \bibinfo{person}{Lovekesh Vig}, {and} \bibinfo{person}{Gautam Shroff}.} \bibinfo{year}{2019}\natexlab{}.
\newblock \showarticletitle{NISER: Normalized Item and Session Representations to Handle Popularity Bias}.
\newblock  (\bibinfo{date}{9} \bibinfo{year}{2019}).
\newblock
\urldef\tempurl%
\url{http://arxiv.org/abs/1909.04276}
\showURL{%
\tempurl}


\bibitem[Haque et~al\mbox{.}(2022)]%
        {TAGNN}
\bibfield{author}{\bibinfo{person}{Sakib Haque}, \bibinfo{person}{Zachary Eberhart}, \bibinfo{person}{Aakash Bansal}, {and} \bibinfo{person}{Collin McMillan}.} \bibinfo{year}{2022}\natexlab{}.
\newblock \showarticletitle{Semantic Similarity Metrics for Evaluating Source Code Summarization}.
\newblock \bibinfo{journal}{\emph{IEEE International Conference on Program Comprehension}}  \bibinfo{volume}{2022-March}, \bibinfo{pages}{36--47}.
\newblock
\showISBNx{9781450392983}
\urldef\tempurl%
\url{https://doi.org/10.1145/nnnnnnn.nnnnnnn}
\showDOI{\tempurl}


\bibitem[Hidasi et~al\mbox{.}(2015)]%
        {GRU4REC}
\bibfield{author}{\bibinfo{person}{Balázs Hidasi}, \bibinfo{person}{Alexandros Karatzoglou}, \bibinfo{person}{Linas Baltrunas}, {and} \bibinfo{person}{Domonkos Tikk}.} \bibinfo{year}{2015}\natexlab{}.
\newblock \showarticletitle{Session-based Recommendations with Recurrent Neural Networks}.
\newblock  (\bibinfo{date}{11} \bibinfo{year}{2015}).
\newblock
\urldef\tempurl%
\url{http://arxiv.org/abs/1511.06939}
\showURL{%
\tempurl}


\bibitem[Jin et~al\mbox{.}(2023)]%
        {NirGNN}
\bibfield{author}{\bibinfo{person}{Di Jin}, \bibinfo{person}{Luzhi Wang}, \bibinfo{person}{Yizhen Zheng}, \bibinfo{person}{Guojie Song}, \bibinfo{person}{Fei Jiang}, \bibinfo{person}{Xiang Li}, \bibinfo{person}{Wei Lin}, {and} \bibinfo{person}{Shirui Pan}.} \bibinfo{year}{2023}\natexlab{}.
\newblock \showarticletitle{Dual Intent Enhanced Graph Neural Network for Session-based New Item Recommendation}.
\newblock \bibinfo{journal}{\emph{ACM Web Conference 2023 - Proceedings of the World Wide Web Conference, WWW 2023}}, \bibinfo{pages}{684--693}.
\newblock
\showISBNx{9781450394161}
\urldef\tempurl%
\url{https://doi.org/10.1145/3543507.3583526}
\showDOI{\tempurl}


\bibitem[Kang and McAuley(2018)]%
        {SASREC}
\bibfield{author}{\bibinfo{person}{Wang-Cheng Kang} {and} \bibinfo{person}{Julian McAuley}.} \bibinfo{year}{2018}\natexlab{}.
\newblock \showarticletitle{Self-Attentive Sequential Recommendation}.
\newblock  (\bibinfo{date}{8} \bibinfo{year}{2018}).
\newblock
\urldef\tempurl%
\url{http://arxiv.org/abs/1808.09781}
\showURL{%
\tempurl}


\bibitem[Landia et~al\mbox{.}(2022)]%
        {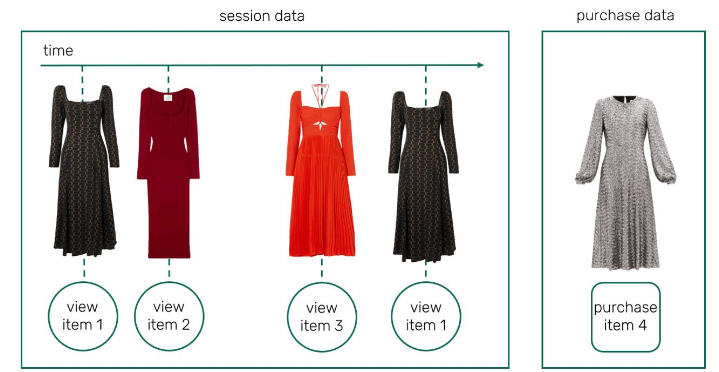}
\bibfield{author}{\bibinfo{person}{Nick Landia}, \bibinfo{person}{Frederick Cheung}, {and} \bibinfo{person}{Donna North}.} \bibinfo{year}{2022}\natexlab{}.
\newblock \bibinfo{title}{Dressipi Datasets}.
\newblock
\newblock
\urldef\tempurl%
\url{https://dressipi.com/datasets/}
\showURL{%
\tempurl}


\bibitem[Li et~al\mbox{.}(2017)]%
        {NARM}
\bibfield{author}{\bibinfo{person}{Jing Li}, \bibinfo{person}{Pengjie Ren}, \bibinfo{person}{Zhumin Chen}, \bibinfo{person}{Zhaochun Ren}, \bibinfo{person}{Tao Lian}, {and} \bibinfo{person}{Jun Ma}.} \bibinfo{year}{2017}\natexlab{}.
\newblock \showarticletitle{Neural attentive session-based recommendation}.
\newblock \bibinfo{journal}{\emph{International Conference on Information and Knowledge Management, Proceedings}}  \bibinfo{volume}{Part F131841}, \bibinfo{pages}{1419--1428}.
\newblock
\showISBNx{9781450349185}
\urldef\tempurl%
\url{https://doi.org/10.1145/3132847.3132926}
\showDOI{\tempurl}


\bibitem[Mitheran et~al\mbox{.}(2021)]%
        {TAGNN++}
\bibfield{author}{\bibinfo{person}{Sai Mitheran}, \bibinfo{person}{Abhinav Java}, \bibinfo{person}{Surya~Kant Sahu}, {and} \bibinfo{person}{Arshad Shaikh}.} \bibinfo{year}{2021}\natexlab{}.
\newblock \showarticletitle{Introducing Self-Attention to Target Attentive Graph Neural Networks}.
\newblock  (\bibinfo{date}{7} \bibinfo{year}{2021}).
\newblock
\urldef\tempurl%
\url{http://arxiv.org/abs/2107.01516}
\showURL{%
\tempurl}


\bibitem[Moreira et~al\mbox{.}(2021)]%
        {Transformer4Rec}
\bibfield{author}{\bibinfo{person}{Gabriel De Souza~Pereira Moreira}, \bibinfo{person}{Sara Rabhi}, \bibinfo{person}{Jeong~Min Lee}, \bibinfo{person}{Ronay Ak}, {and} \bibinfo{person}{Even Oldridge}.} \bibinfo{year}{2021}\natexlab{}.
\newblock \showarticletitle{Transformers4Rec: Bridging the Gap between NLP and sequential/session-based recommendation}.
\newblock \bibinfo{journal}{\emph{RecSys 2021 - 15th ACM Conference on Recommender Systems}}, \bibinfo{pages}{143--153}.
\newblock
\showISBNx{9781450384582}
\urldef\tempurl%
\url{https://doi.org/10.1145/3460231.3474255}
\showDOI{\tempurl}


\bibitem[Sun et~al\mbox{.}(2019)]%
        {BERT4REC}
\bibfield{author}{\bibinfo{person}{Fei Sun}, \bibinfo{person}{Jun Liu}, \bibinfo{person}{Jian Wu}, \bibinfo{person}{Changhua Pei}, \bibinfo{person}{Xiao Lin}, \bibinfo{person}{Wenwu Ou}, {and} \bibinfo{person}{Peng Jiang}.} \bibinfo{year}{2019}\natexlab{}.
\newblock \showarticletitle{Bert4rec: Sequential recommendation with bidirectional encoder representations from transformer}.
\newblock \bibinfo{journal}{\emph{International Conference on Information and Knowledge Management, Proceedings}}, \bibinfo{pages}{1441--1450}.
\newblock
\showISBNx{9781450369763}
\urldef\tempurl%
\url{https://doi.org/10.1145/3357384.3357895}
\showDOI{\tempurl}


\bibitem[Wu et~al\mbox{.}(2018)]%
        {SRGNN}
\bibfield{author}{\bibinfo{person}{Shu Wu}, \bibinfo{person}{Yuyuan Tang}, \bibinfo{person}{Yanqiao Zhu}, \bibinfo{person}{Liang Wang}, \bibinfo{person}{Xing Xie}, {and} \bibinfo{person}{Tieniu Tan}.} \bibinfo{year}{2018}\natexlab{}.
\newblock \showarticletitle{Session-based Recommendation with Graph Neural Networks}.
\newblock  (\bibinfo{date}{10} \bibinfo{year}{2018}).
\newblock
\urldef\tempurl%
\url{https://doi.org/10.1609/aaai.v33i01.3301346}
\showDOI{\tempurl}


\bibitem[Xie et~al\mbox{.}(2022)]%
        {DIF-SR}
\bibfield{author}{\bibinfo{person}{Yueqi Xie}, \bibinfo{person}{Peilin Zhou}, {and} \bibinfo{person}{Sunghun Kim}.} \bibinfo{year}{2022}\natexlab{}.
\newblock \showarticletitle{Decoupled Side Information Fusion for Sequential Recommendation}.
\newblock \bibinfo{journal}{\emph{SIGIR 2022 - Proceedings of the 45th International ACM SIGIR Conference on Research and Development in Information Retrieval}}, \bibinfo{pages}{1611--1621}.
\newblock
\showISBNx{9781450387323}
\urldef\tempurl%
\url{https://doi.org/10.1145/3477495.3531963}
\showDOI{\tempurl}


\bibitem[Xue et~al\mbox{.}(2022)]%
        {SIHG4SR}
\bibfield{author}{\bibinfo{person}{Chendi Xue}, \bibinfo{person}{Xinyao Wang}, \bibinfo{person}{Yu Zhou}, \bibinfo{person}{Ke Ding}, \bibinfo{person}{Jian Zhang}, \bibinfo{person}{Rita~Brugarolas Brufau}, {and} \bibinfo{person}{Eric Anderson}.} \bibinfo{year}{2022}\natexlab{}.
\newblock \showarticletitle{SIHG4SR: Side Information Heterogeneous Graph for Session Recommender}.
\newblock \bibinfo{journal}{\emph{ACM International Conference Proceeding Series}}, \bibinfo{pages}{55--63}.
\newblock
\showISBNx{9781450398565}
\urldef\tempurl%
\url{https://doi.org/10.1145/3556702.3556852}
\showDOI{\tempurl}


\bibitem[Zhang et~al\mbox{.}(2019)]%
        {FDSA}
\bibfield{author}{\bibinfo{person}{Tingting Zhang}, \bibinfo{person}{Pengpeng Zhao}, \bibinfo{person}{Yanchi Liu}, \bibinfo{person}{Victor~S Sheng}, \bibinfo{person}{Jiajie Xu}, \bibinfo{person}{Deqing Wang}, \bibinfo{person}{Guanfeng Liu}, {and} \bibinfo{person}{Xiaofang Zhou}.} \bibinfo{year}{2019}\natexlab{}.
\newblock \bibinfo{title}{Feature-level Deeper Self-Attention Network for Sequential Recommendation}.
\newblock
\newblock


\bibitem[Zhang et~al\mbox{.}(2022)]%
        {CoHHN}
\bibfield{author}{\bibinfo{person}{Xiaokun Zhang}, \bibinfo{person}{Bo Xu}, \bibinfo{person}{Liang Yang}, \bibinfo{person}{Chenliang Li}, \bibinfo{person}{Fenglong Ma}, \bibinfo{person}{Haifeng Liu}, {and} \bibinfo{person}{Hongfei Lin}.} \bibinfo{year}{2022}\natexlab{}.
\newblock \showarticletitle{Price DOES Matter!: Modeling Price and Interest Preferences in Session-based Recommendation}.
\newblock \bibinfo{journal}{\emph{SIGIR 2022 - Proceedings of the 45th International ACM SIGIR Conference on Research and Development in Information Retrieval}}, \bibinfo{pages}{1684--1693}.
\newblock
\showISBNx{9781450387323}
\urldef\tempurl%
\url{https://doi.org/10.1145/3477495.3532043}
\showDOI{\tempurl}


\end{thebibliography}

\appendix

\end{document}